\def\im{{\Im\rm m}\mathop}
\documentclass[journal=nalefd,manuscript=letter
]{achemso}
\setkeys{acs}{
hyperref=true}

\usepackage{url}
\usepackage{graphicx}
\usepackage{revsymb}
\usepackage{color}
\usepackage{hyperref}
\makeatletter
\def\Hy@safe@activestrue{}
\makeatother
\title{Continuous third harmonic generation in a terahertz driven modulated
  nanowire}
\author{Kathleen E. Hamilton}
\email{kathleen.hamilton@email.ucr.edu}
\author{Alexey A. Kovalev}
\author{Amrit De}
\author{Leonid P. Pryadko}
\affiliation{Department of Physics \& Astronomy, University of California, 
Riverside, California, 92521, USA}

\begin{document}

\begin{abstract}
  We consider the possibility of observing continuous third-harmonic
  generation using a strongly driven, single-band one-dimensional
  metal. In the absence of scattering, the quantum efficiency of
  frequency tripling for such a system can be as high as 93\%.
  Combining the Floquet quasi-energy spectrum with the Keldysh Green's
  function technique, we derive the semiclassical master equation for
  a one-dimensional band of strongly and rapidly driven electrons in
  the presence of weak scattering by phonons.  The power absorbed from
  the driving field is continuously dissipated by phonon modes,
  leading to a quasi-equilibrium in the electron distribution.  We use
  the Kronig-Penney model with varying effective mass to establish
  growth parameters of an InAs/InP nanowire near optimal for 
third harmonic generation at terahertz frequency range.
\end{abstract}

\maketitle

\section{Introduction}
\label{sec:Introduction}
When electrons in a crystal band are driven by an external
time-independent electric field, they move periodically across the
Brillouin zone, creating characteristic Bloch
oscillations\cite{Bloch-1928,Yakovlev-FTT-1961,%
  Yakovlev-JETP-1961,Keldysh-1963,Wannier-1962}.  The frequency of 
the
oscillations, $\omega_\mathrm{B}=eE a/\hbar$, where $a$ is the unit
cell size, coincides with the energy separation between neighboring
states localized on a Wannier-Stark ladder
\cite{Wannier-1962,Fukuyama-1973}.  The effect has been observed for
electrons/holes in semiconducting
superlattices\cite{Waschke-1993,Mendez-Bastard-1993}, for atoms
trapped in a periodic optical potential\cite{BenDahan-1996}, and for
light propagating in a periodic array of waveguides, with gradient of
the temperature or of the refraction index working as an effective
electric field\cite{Pertsch-1999,Morandotti-1999,Demetrios-2003}.

Combining the effects of a strong, time-periodic driving field, with the nonlinearity of the Bloch oscillations leads to higher harmonic generation of the driving frequency\cite{Faisal-Kaminski-1997,Gupta-Alon-Moiseyev-2003,Golde-Meier-Koch-2008}.  This effect has recently been observed in
bulk ZnO crystals strongly driven by a few-cycle pulsed infrared
laser\cite{Ghimire-2011}.   The application of the infrared field in short, 100-femtosecond, pulses was necessary to ensure that the absorbed energy could be transferred to the lattice and dissipated.

\begin{figure}[htbp]
 \centering
\includegraphics[width=3in
]{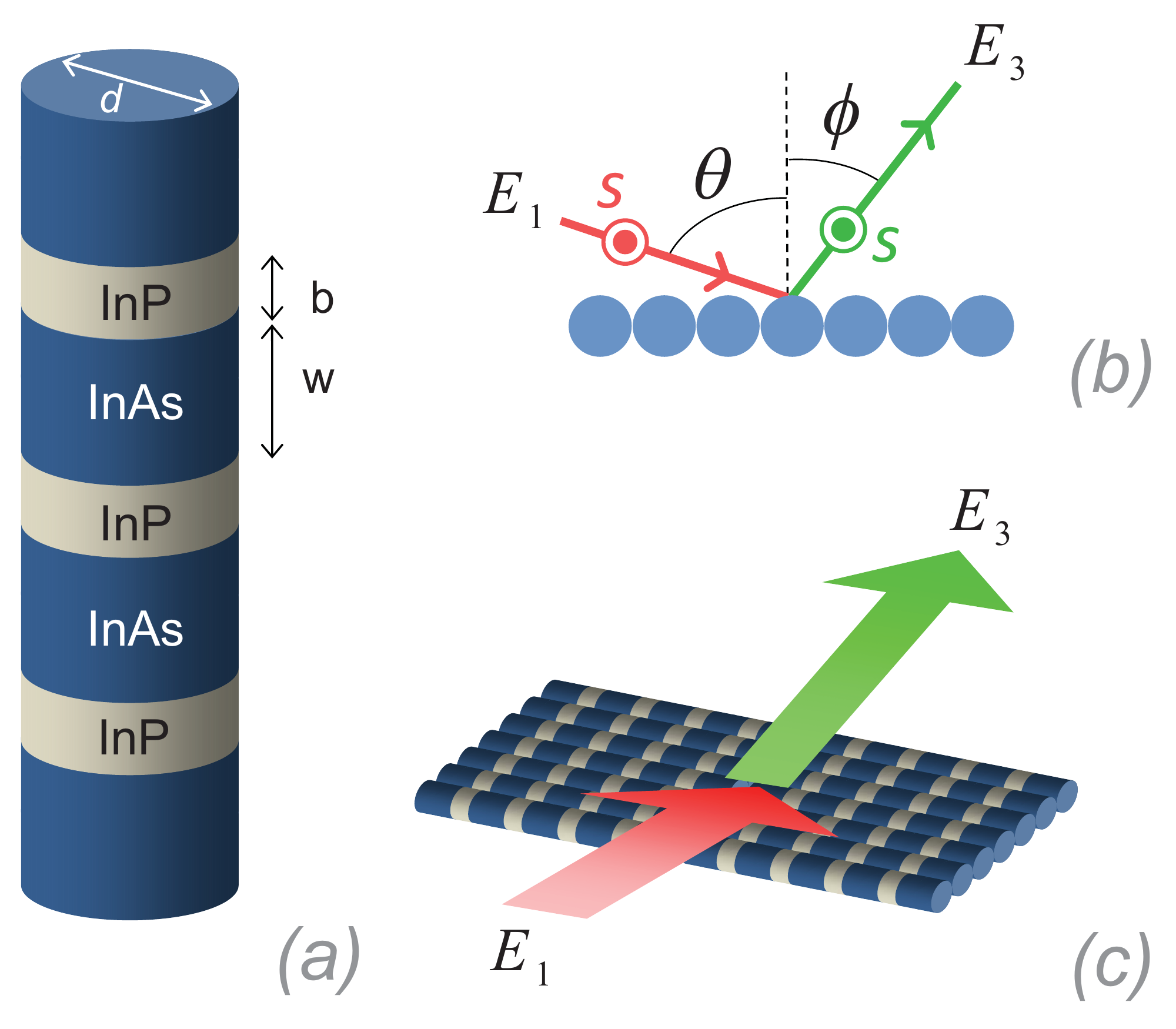}
\caption{(a) A nanowire made with alternating InAs/InP
  regions. (b), (c) Schematic of the third harmonic generation with a
  planar array of such nanowires.  The driving field is $s$-polarized
  so that the electric field $E_1$ be parallel to the nanowires.  The
  generated third harmonic will have the same polarization but
  propagate at a different angle.}\label{fig:nanowire}
\end{figure}

In this work, we suggest that frequency multiplication due to periodically-driven Bloch oscillations could also be observed in a steady-state setting, e.g., a periodically modulated nanowire (or an array of such nanowires) continuously driven by high-amplitude terahertz radiation (see Fig.~\ref{fig:nanowire}). In the weak-scattering limit, the quantum efficiency of frequency tripling for such a system can be as high as 93\%. 

For a nanowire in mechanical contact with an insulating, optically 
transparent substrate, a quasi-equilibrium electron distribution will be 
reached as the power absorbed from the driving field will be continuously 
dissipated into phonon modes. This distribution can be quite different from 
the initial, equilibrium Fermi distribution. In particular, at the driving field 
amplitude which is optimal for third harmonic generation, the distribution 
can be both broadened and inverted.  The inversion of the distribution 
occurs once the driving field amplitude exceeds the dynamical localization 
threshold\cite{Dunlap-Kenkre-1986,
Grossmann-Dittrich-Jung-Hanggi-1991}.

In our analytic derivation, we combine the Floquet quasi-energy description with the
Keldysh Green's function technique to obtain the semiclassical master
equation for a one-dimensional band of strongly and rapidly driven
electrons in the presence of weak scattering by phonons.  We solve
these equations numerically to find the electron distribution function for a
cosine energy band at a given driving field frequency (fixed at
$\Omega/2\pi=1$\,THz) and the field amplitude chosen to suppress the
generation of the principal harmonic.  This electron distribution is
used as an input for calculating the time-dependent current and the
intensity radiated at different harmonics of the driving field
frequency.  We use these results 
to find the optimal dimensions
of a periodically modulated InAs/InP nanowire, which would yield the most efficient
frequency tripling of $1$\,THz radiation.


\section{Theoretical approach}

We consider a single-band one-dimensional metallic wire driven by a 
harmonic
electric field with the amplitude $E_0$ and frequency $\Omega$, and
coupled to substrate phonons,
\begin{equation}
  H=H_0+H_\mathrm{e-ph}+H_\mathrm{ph},
\end{equation}
where the electron, electron-phonon, and phonon Hamiltonians are, 
respectively
\begin{eqnarray}
  \label{eq:ham-electron}
  H_0&=&\sum_k \varepsilon\biglb(k+A(t)\bigrb) c_{k}^\dagger c_{k},\\ 
  \label{eq:ham-e-ph}
  H_\mathrm{e-ph}&=&V^{-1/2}\sum_{\mathbf{q},k} 
  { M_{{\bf q},k}}
  c_{k+q_\parallel}^\dagger c_k (b_{\bf q}+b_{-\bf q}^\dagger) , \\
  \label{eq:ham-phonon}
  H_{\rm ph}&=&\sum_{\bf q}\omega_{\bf q} b_{\bf q}^\dagger b_{\bf 
q}.
\end{eqnarray}
Here $c_k$ $(c_{k}^\dagger)$ is the annihilation (creation) operator
for an electron with one-dimensional momentum $\hbar k$ and energy
$\varepsilon(k)$. To apply our results to a periodically modulated 
nanowire,  we assume a tight-binding model with the electronic spectrum,
 \begin{equation}
\varepsilon(k)=-2J \cos (ka),\label{eq:tight-binding}
\end{equation}
where $J$ is the hopping matrix element and $a$ is the period of the
potential along the chain.
The electric field is incorporated into the Hamiltonian through the vector 
potential 
$A(t)=A_0 \sin\Omega t$ with $A_0=eE_0/\hbar \Omega$ representing 
the
vector potential of the driving field. Phonon annihilation (creation) 
operators $b_{\mathbf{q}}$ and $b_{ \mathbf{q}}^\dagger$  are labeled 
with the three-dimensional wavevector  $\mathbf{q}\equiv
(q_\parallel,\mathbf{q}_\perp)$ and
$\omega_{\mathbf{q}}$ is the phonon frequency (electron spin and 
phonon branch
indices are suppressed). The factors $M_{{\bf
    q},k}=\alpha_{\mathbf{q},k}(\hbar/2\omega_{\mathbf{q}})^{1/2}$ are 
the
matrix elements for electron-phonon scattering.

We ignore the effects of disorder or electron-electron interactions,
and consider lattice phonons in thermal equilibrium at temperature
$\hbar/k_B\beta$.  We do not include directly the scattering by phonon
modes of the nanowire, assuming that they are strongly hybridized with
those of the substrate, with the corresponding effects  incorporated
in the matrix elements $M_{\mathbf{q},k}$. The electron-phonon
coupling is considered to be weak, meaning that the phonon scattering
time is long compared to the period $\tau\equiv 2\pi/\Omega$ of the
driving field.

\subsection{Modified energy spectrum of the driven system}
The dynamics of the strongly-driven electrons with the Hamiltonian
(\ref{eq:ham-electron}) is characterized by non-monotonous phases
\begin{equation}
  \varphi_{k}(t) = \int_0^{t} dt^{\prime} \varepsilon\biglb(k +
  A(t^{\prime})\bigrb). 
  \label{eq:phases}
\end{equation}
The phase accumulated over a period,
$ \varphi_{k}(\tau)$, can be
expressed in terms of the average
particle 
energy with the momentum $\hbar k$,
\begin{equation}\label{eq:average-energy}
\langle  \varepsilon(k +
A)\rangle\equiv {\tau}^{-1}\int_0^\tau dt\,\varepsilon\biglb(k +
A(t)\bigrb);
\end{equation}
clearly, this energy can be also identified as the Floquet energy of a
single-electron state.  While Eq.~(\ref{eq:average-energy}) does not
include the usual additive uncertainty $m\Omega$, this particular
choice has the advantage that in the weak-field limit, $A_0\to0$,
$\langle \varepsilon(k + A)\rangle$ recovers the zero-field spectrum
$\varepsilon(k)$.  

The average energy~(\ref{eq:average-energy}) also coincides with that
introduced in the theory of dynamical
localization\cite{Dunlap-Kenkre-1986,
Grossmann-Dittrich-Jung-Hanggi-1991}.
Dynamical localization occurs when the effective band becomes
flat, i.e., $\langle \varepsilon(k+A)\rangle \to0$.  The corresponding
condition is most easily obtained in the special case of tight-binding
model with the spectrum~(\ref{eq:tight-binding}),
\begin{equation}
  \label{eq:tight-binding-average}
  \langle \varepsilon(k+A)\rangle=-2\widetilde J \cos (k a),\quad \widetilde
  J\equiv J \,J_0(A_0 a), 
\end{equation}
where $J_0(z)$ is the zeroth order Bessel function.   With
the driving field amplitude increasing from zero the bandwidth is
gradually reduced; it switches sign at the roots of the Bessel
function, $A_0\, a=\zeta_{0n}$.  The first time this happens corresponds
to the electric field $E_0=\zeta_{01}\hbar \Omega/ea$, where
$\zeta_{01}\approx 2.405$.

\subsection{Frequency Multiplication with weak scattering}
\label{sec:Frequency_Multiplication}
We obtain the instantaneous current by averaging the canonical
velocity operator $\partial H/\partial A$ over the electron distribution
function $f_k\equiv \langle c_k^\dagger
c_k\rangle$, 
\begin{equation}
i(t) =  C_f(t)\,\sin A(t) a+ S_f(t)\,\cos A(t) a, 
\end{equation}
where we assumed the tight-binding spectrum~(\ref{eq:tight-binding})
and used the definitions 
\begin{equation}
  \label{eq:params}
  C_f(t)\equiv  2J \int \frac{dk}{2\pi}  \cos(k) f_k,\quad 
  S_f(t)\equiv  2J \int \frac{dk}{2\pi}  \sin(k) f_k.
\end{equation}

\begin{figure}[h]
 \centering
 \includegraphics[width=3in]{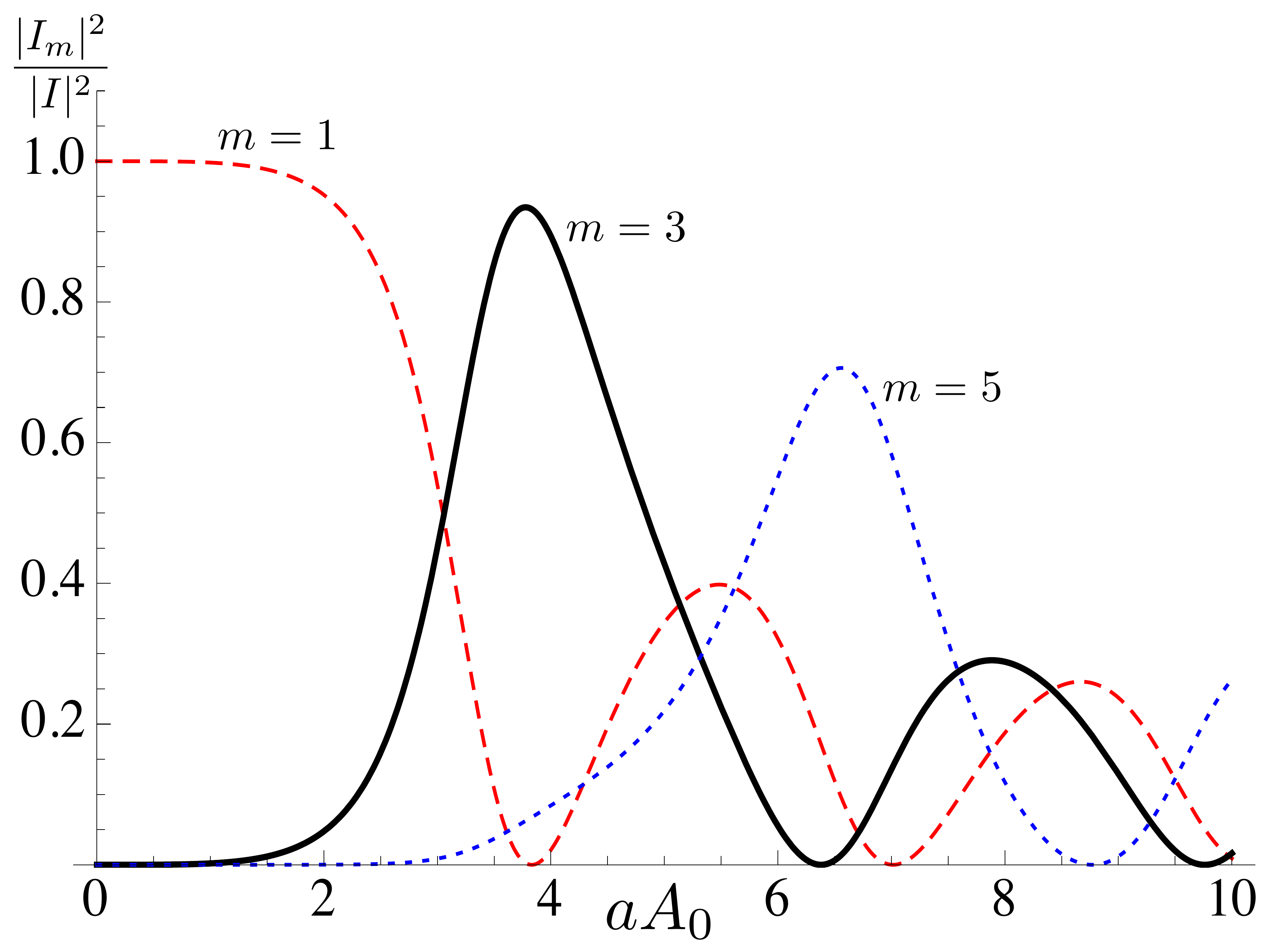}
 \caption{Normalized magnitude squared of the Fourier harmonics of the
   instantaneous current, $|I_m|^2$, for $m=1$ (red dashed), $m = 3$
   (black, solid), and $m = 5$ (blue, dotted) plotted as a function of
   the dimensionless amplitude of the vector potential of the driving
   field, see Eq.~(\ref{eq:ham-electron}).  The intensities $|I_m|^2$
   correspond to the power emitted in the corresponding harmonics when
   multiple nanowires are used in a planar geometry, see
   Fig.~\ref{fig:nanowire} (b),(c).}
\label{fig:AnalyticHarmonics}
\end{figure}

In the limit of weak scattering, the distribution function $f_k$ is time-
independent and 
always symmetric,
$f_{-k}=f_k$.  Thus, $S_f(t)=0$ while $C_f(t)=C_f$ is a 
time-independent pre-factor. The Fourier components of the current are
obtained directly,
\begin{equation}
i(t)= 
  2 C_f \sum_{m=1,3,5,\ldots}  J_m(A_0 a) \sin(m\Omega t), 
\end{equation}
where the summation is over the odd harmonics $m$.  By choosing $A_0a
= \zeta_{11}\approx 3.8317$, the first harmonic can be fully
suppressed, which leaves the third harmonic dominant.  The maximal 
value for the fraction of
the energy emitted into the third harmonic ($93.34\%$) is found in close 
vicinity of this amplitude, see
Fig.~\ref{fig:AnalyticHarmonics}.


\subsection{Transition kinetics in  a driven system}
\label{sec:qke}


We use the Keldysh non-equilibrium Green's function (GF)
formalism\cite{Keldysh64,Rammer-Smith-1986,%
  Kamenev-2004,Haug-Jauho-2008}
along with a perturbation theory expansion with respect to the entire
time-dependent electron Hamiltonian~(\ref{eq:ham-electron}); the
corresponding evolution is solved exactly in terms of the
phases~(\ref{eq:phases}).  Previously, related approaches have been
used, e.g., for describing ionization of
atoms\cite{DeVries-1990,Joachain-book-2011} and the high-order
harmonic generation\cite{Kemper-Moritz-Freericks-Devereaux-2013} in
the field of ultrashort laser pulses.  Here, instead of solving the
corresponding equations numerically, we take the limit of weak
electron-phonon coupling and analytically derive the semiclassical
master equation for electron distribution function averaged over the
period of the driving field, see Eqs.~(\ref{eq:qke}) and
(\ref{eq:rates}).  The same master equation can also be derived from the
formalism by Konstantinov and Perel'\cite{Konstantinov-Perel-1960}
with the help of an appropriate resummation of the perturbation
series\cite{pryadko-sengupta-kinetics-2006}.

In the interaction representation with respect
to the time-dependent Hamiltonian~(\ref{eq:ham-electron}), the
electron operators acquire time-dependence $ e^{-i\varphi_k(t)}c_k$
with quasiperiodic phases~(\ref{eq:phases}).  We separate these phases
by defining the ``lower-case'' GFs 
\begin{equation}
  \label{eq:gf}
g_k(t_2,t_1) = e^{-i\varphi_k(t_2)}   G_k(t_2,t_1) e^{i\varphi_k (t_1)},
\end{equation}
where the ``upper-case'' $G_k(t_2,t_1)$ is any of the conventional GFs
introduced in the Keldysh formalism\cite{Keldysh64,
Rammer-Smith-1986,%
  Kamenev-2004,Haug-Jauho-2008}.  These phases introduce rapid
oscillations in the self-energy, making the direct Wigner
transformation difficult.  We notice, however, that in the limit of weak
electron-phonon coupling, the GFs (\ref{eq:gf}) are expected to change
only weakly when both time arguments are incremented by the driving
period $\tau$.  This implies that in the following decomposition,
\begin{equation}
  \label{eq:gf-expanded}
  g_k(t_2,t_1)=\sum_m g_{k,m}(t,T)e^{-im\Omega T},
\end{equation}
$t\equiv t_2-t_1$ is the ``fast'' time, while $T\equiv (t_2+t_1)/2$ is
the ``slow'' time when it appears as an argument of thus defined
Floquet components $g_{k,m}(t,T)$ of the GF.  The Dyson equations for 
thus defined 
Keldysh $g_{k,m}^K$ and retarded $g_{k,m}^R$
GFs\cite{Rammer-Smith-1986} have the form
\begin{eqnarray}
(  i\partial_T +m\Omega) g_{k,m}^K(t,T)&=&I^K_\mathrm{coll},\\
  i\partial_t g_{k,m}^R(t,T)&=&\delta_{m,0}\,\delta(t)+I^R_\mathrm
{coll},
\end{eqnarray}
where $I^K_\mathrm{coll}$ and $I^R_\mathrm{coll}$ are the collision
integrals originating from the corresponding self-energy functions.
The collision integrals being relatively small, both $g_{k,m}^K$ and
$g_{k,m}^R$ are dominated by the $m=0$
components\cite{Hamilton-Kovalev-Pryadko-2012}.

To derive the semiclassical master equation, we write the equations
for the $m=0$ components of the ``lesser'' $g^<$ and ``greater'' $g^>$
GFs\cite{Rammer-Smith-1986}, perform the Wigner transformation
replacing the fast time variable $t$ by the frequency
$\omega$, and use a version of the Kadanoff-Baym
approximation\cite{Kadanoff-Baym-book-1962}
\begin{equation}
  g^{<}_{k,0}(\omega,T) = i A_{k,0}(\omega,T) f_k(T),\quad 
  A_{k,0}(\omega)  \approx  \delta(\omega),
\label{eq:KB}
\end{equation}
 for the corresponding
spectral function,
$A_{k,0}(\omega,T)=i[g^<_{k,0}(\omega,T)-g^>_{k,0}(\omega,T)]=2\,
\im
g^R_{k,0}(\omega,T)$,
where $f_k(T)$ is the non-equilibrium electron distribution function
averaged over the period.  This requires that the electron-pho\-non
coupling be weak, and assumes that the electron spectrum renormalization 
has
been included in the Hamiltonian (\ref{eq:ham-electron}).

The resulting master equation for weak electron-pho\-non interactions has
the following standard form\cite{Hamilton-Kovalev-Pryadko-2012}
\begin{eqnarray}
  \nonumber 
  \frac{d}{ dt}{f}_k(T) &=& \int \frac{dk'}{ 2\pi} \Bigl\{\Gamma_{k,k'}\,
[ 1 -
    f_{k'}(T) ] f_k(T) \\  & & \qquad\;\; 
  -\Gamma_{k',k}\,f_{k'}(T)[ 1 - f_{k}(T) ]\Bigr\},
  \label{eq:qke}
\end{eqnarray}
where the transition rates are
\begin{eqnarray}
  \nonumber 
  \Gamma_{k,k'}&=& 2 \sum_m |S_{k,k'}(m)|^2 \int_0^\infty\!\!\! {d
\omega}\,
  W_{k,k'}(\omega)\\
  & &\hskip-0.45in \times \left[(n_\omega + 1)
    \delta(\Delta\varepsilon_{k,k'}^{(m)}-\hbar\omega) +  n_\omega
    \delta(\Delta\varepsilon_{k,k'}^{(m)}+
\hbar\omega)\right].\;\;\strut
\label{eq:rates}
\end{eqnarray}
Here $W_{k,k'}(\omega)$ is the phonon spectral function (density of
states weighted by the square of the coupling) for a given momentum
$q_\parallel =k'-k$ along the wire, see Eq.~(\ref{eq:ham-e-ph}),
$n_\omega\equiv [\exp(\beta\omega)-1]^{-1}$ is the phonon distribution
function, and the energy increment
\begin{equation}
  \label{eq:energy-increment}
  \Delta\varepsilon_{k,k'}^{(m)}\equiv
\langle\varepsilon(k+A)\rangle-\langle\varepsilon(k'+A)\rangle-m\,\hbar
\Omega, 
\end{equation}
is the energy carried in or out by phonons, depending on its sign.
Note that this energy includes $m$ quanta of the driving field,
emitted or absorbed, depending on the sign of $m=0,\pm1,\ldots$.  The
matrix elements $S_{k,k'}(m) $ are the Fourier expansion
coefficients of the product of the two phase factors,
$e^{i\delta\varphi_k(t)-i\delta\varphi_{k'}(t)}$, where
$\delta\varphi_k(t)\equiv \varphi_k(t)-t\langle\varepsilon(k+A)\rangle$
is the periodic part of the phase.  They satisfy the sum rule 
\begin{equation}
  \label{eq:sum-rule}
  \sum_{m=-\infty}^\infty |S_{k,k'}(m)|^2=1.
\end{equation}
Clearly, the equilibrium Fermi distribution for $f_k$ is only obtained
in the limit of small electric field amplitudes, such that 
$S_{k,k'}(m)$ with $m=0$ gives the dominant contribution.


\section{Simulation results} 
\label{sec:results}

The results presented in this section have been obtained by
numerically finding the stationary solution of the discretized version
of the master equation~(\ref{eq:qke}) with transition
rates~(\ref{eq:rates}). A simple model for the phonon spectral
function,
$W_{k,k'}(\omega)=\gamma^2\,\theta(\omega-s\left|k-k'\right|)$, was 
used, with
the sound speed $s=5\times 10^3$\,m/s as appropriate for typical 3D
acoustical phonons.  Since we assume no other scattering mechanisms,
the quasi-equilibrium distribution functions $f_k$ and other results do
not depend on the magnitude of the electron-phonon coupling
$\gamma^2$.

We fix the phonon temperature at $4.2$K, the lattice period
$a=8.64$\,nm, the average electron filling at $1/2$ and choose the
driving field frequency $\Omega/2\pi=10^{12}$ Hz (energy
$\hbar\Omega\approx 4.14$\,meV). Also, the amplitude $A_0
a=\zeta_{11}\approx3.8317$ is fixed, which corresponds to the point 
where the
first harmonic generation is fully suppressed [see
  Fig.~\ref{fig:AnalyticHarmonics}].  At this point the effective
coupling is $\widetilde J=J\,J_0(\zeta_{11})\approx -0.403\, J$, which
creates an inverted and somewhat narrowed band.  The effective
bandwidth is smaller than $\hbar\Omega$ for $J<2.57$\,meV.

\begin{figure}[htbp] \centering
\includegraphics[width=3in]{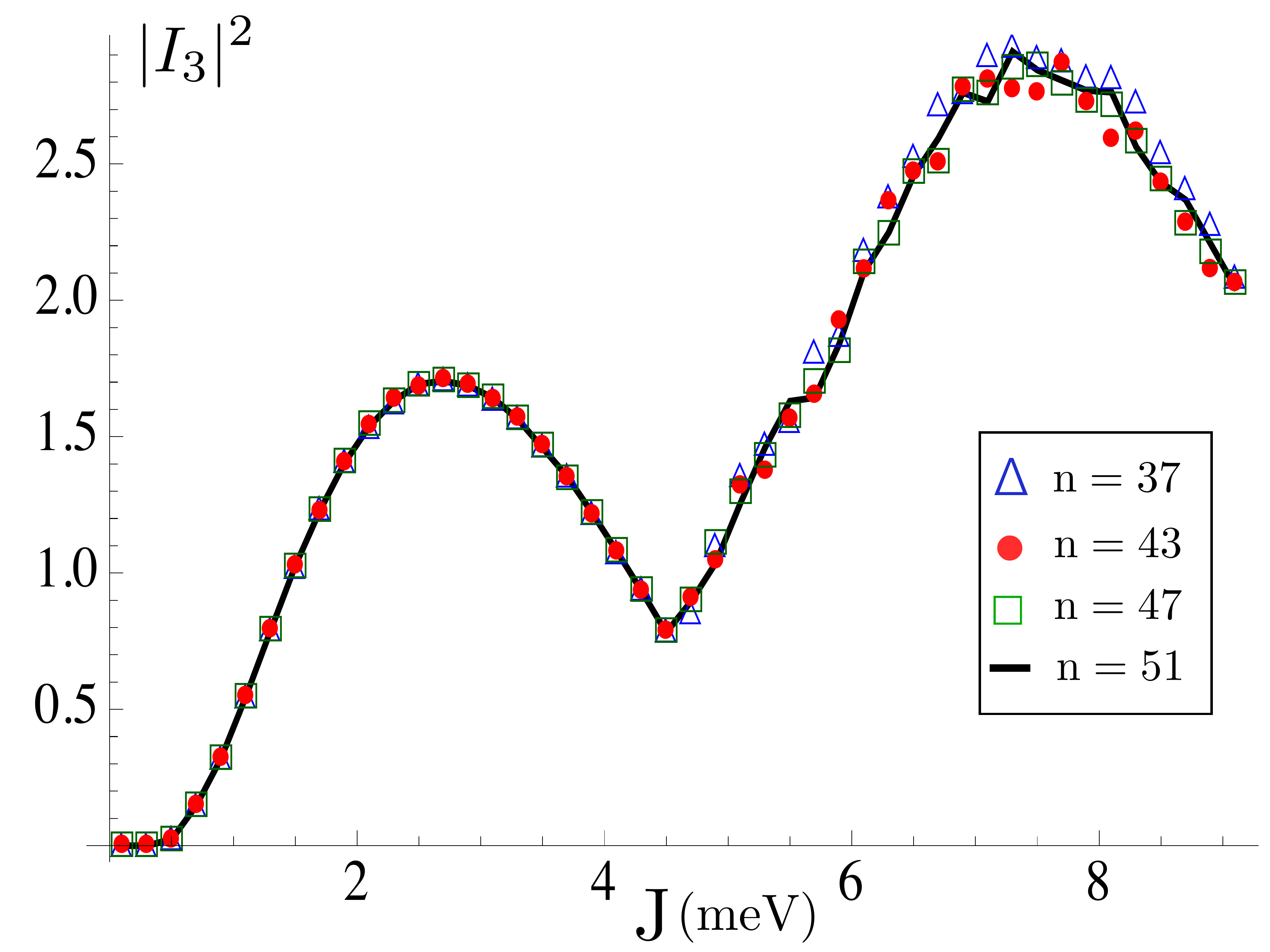}
\caption{Magnitude squared of the third harmonic of the instantaneous
  current (arbitrary units), plotted as a function of the
  tight-binding parameter $J$, computed with $N_k=37,43,47,51$
  discrete momentum points as indicated in the caption.  See text for
  other simulation parameters.  The pronounced minima are caused by
  the flattening of the distribution function near thresholds of
  $m$-photon-assisted scattering between the vicinities of $k=0$ and
  $k=\pm\pi/a$, with $m$ even.}\label{fig:I3vJ}
\end{figure}

\begin{figure}[htbp]
 \centering
\includegraphics[width=2.7in]{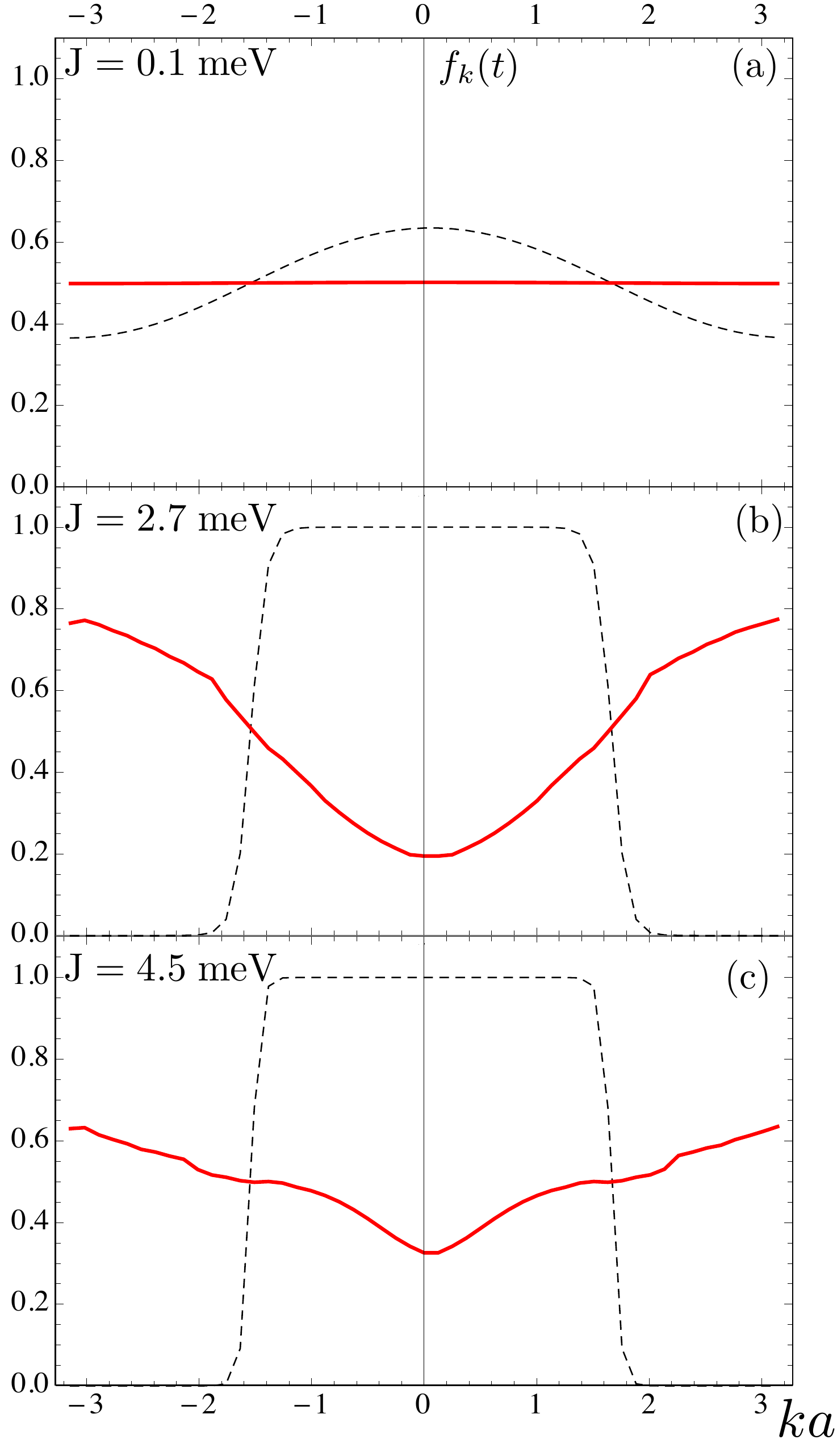}
\caption{Solid lines: the stationary distribution functions obtained by solving
  discretized versions of Eqs.~(\ref{eq:qke}), (\ref{eq:rates}) with
  $N_k=51$ momentum points and  the tight-binding parameters $J$ as
  indicated.   See text for other simulation parameters. Dashed lines: equilibrium
  Fermi distribution functions. 
}\label{fig:3Distributions}
\end{figure}

\begin{figure}[htbp]
 \centering
\includegraphics[width=3in]{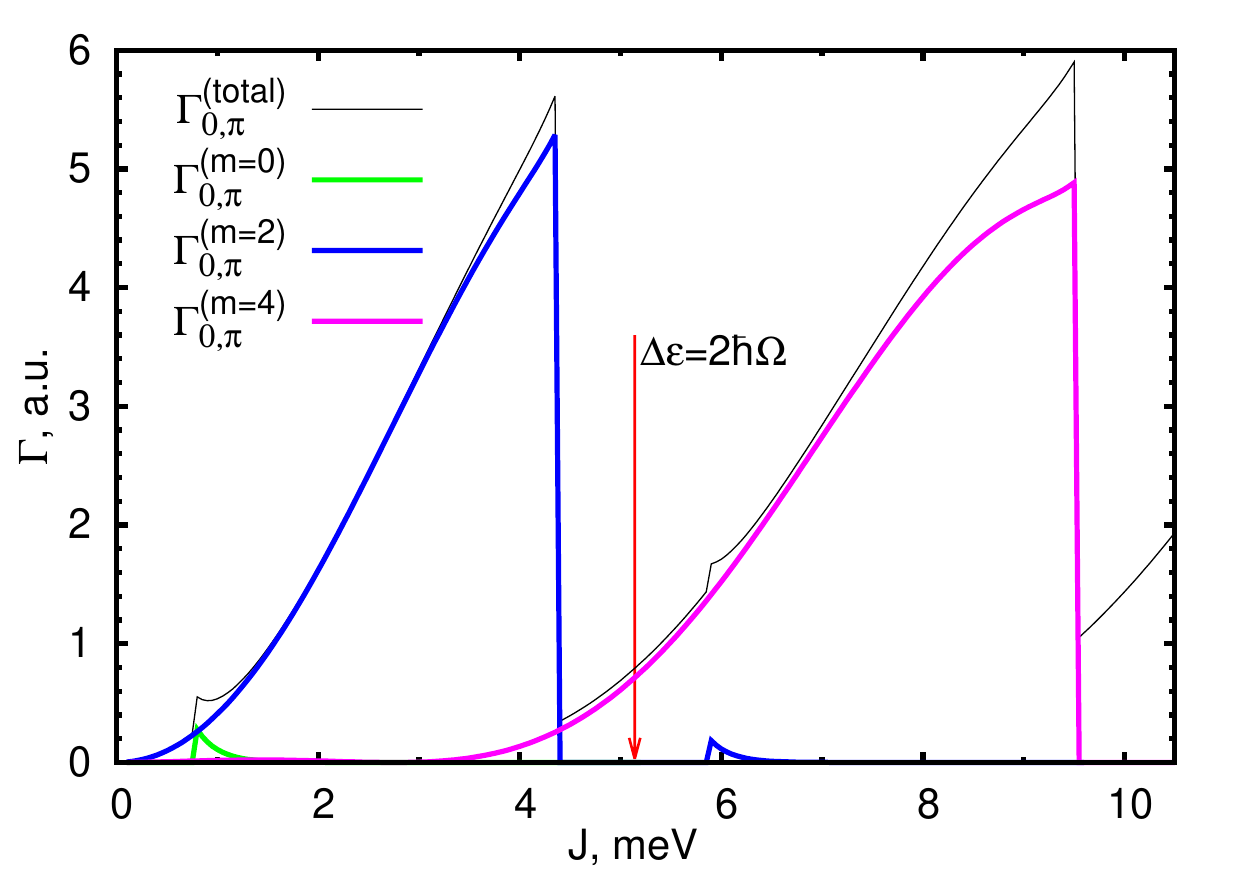}
\caption{The transition rate~(\ref{eq:rates}) (arbitrary units) for
  scattering between the sites at $k = -\pi/a$ and $k = 0$ (black,
  solid) and the individual contributions from $m$-photon assisted
  processes as indicated.  The vertical dashed line at $J=5.1$\,meV
  indicates the threshold for the $m=2$ transition, $|4\widetilde{J}|=2\hbar
\Omega$;
  the peaks to the left and to the right of this point correspond to
  phonon emission and absorption,
  respectively.}\label{fig:GammaRatePlot1}
\end{figure}

In Fig.~\ref{fig:I3vJ}, we show the intensity $|I_3|^2$ of the
radiated third harmonic (in arbitrary units) as a function of the
tight-binding hopping parameter $J$.  The overall upward trend
reflects the linear scaling of the current with $J$.  The plot has a
series of pronounced maxima and minima related to the structure of the
distribution function $f_k$, see Fig.~\ref{fig:3Distributions}.
Indeed, at the first maximum of the radiated intensity $|I_3|^2$,
{$J=2.7$\,meV}, the distribution function has a well-defined minimum
at $k=0$ and symmetric maxima at $k=\pm\pi/a$
[Fig.~\ref{fig:3Distributions}(b)]; notice the population inversion
consistent with negative $\widetilde J$.  On the other hand, the
distribution in Fig.~\ref{fig:3Distributions}(c) corresponding to the
first minimum of radiated intensity, $J=4.5$\,meV, is much flatter.
This flattening can be traced to a sharp increase of the transition
rates connecting the regions of momentum space near $k=0$ and $k=\pi/a
$.
This is illustrated in Fig.~\ref{fig:GammaRatePlot1}, where transition
rates between $k=0$ and $k=\pi/a$ are shown.  The corresponding phases
$\delta\varphi_{\pi/a}=-\delta\varphi_{0}$ have only even harmonics
$m\Omega$, $m=2,4,\ldots$, and the threshold values of $J$ for
different $m$ correspond to sharp maxima of $\Gamma_{0,\pi/a}$.

\begin{figure}[htbp]
 \centering
\includegraphics[width=2.9in]{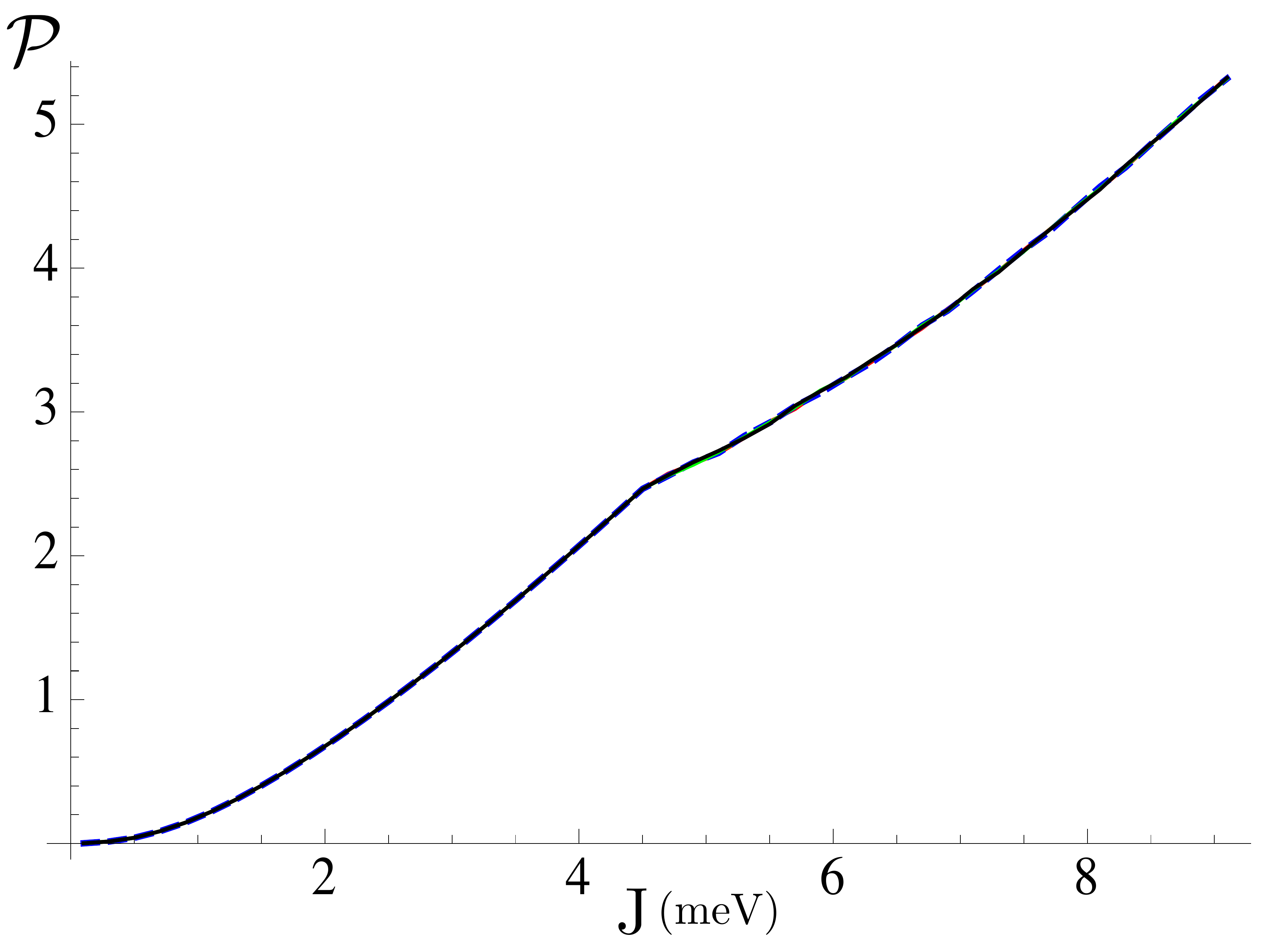}
\caption{Scaling of the average power (arbitrary units) dissipated
  into the phonon modes as a function of the
  tight-binding parameter $J$.  The four curves on top of each other
  correspond to the same numbers of discrete momentum points $N_k$ as
  in Fig.~\ref{fig:I3vJ}.}\label{fig:PowervsJ}
\end{figure}

In Fig.~\ref{fig:PowervsJ}, we show how the average power ${\cal P}$
radiated into the phonon modes scales with the tight-binding parameter
$J$.  While general dependence on $J$ is monotonic, at $J=4.5$\,meV,
where the third harmonic has a minimum, ${\cal P}$ changes slope.

\section{Proposed Experimental Design}
\label{sec:nanowire}
The simulation results in the previous section suggest that the
optimal system for third harmonic generation would be a
one-dimensional metallic conductor with an unrenormalized bandwidth
close to $2.6$ times the energy $\hbar\Omega$ of the driving field
quanta (bandwidth of about $11$\,meV for $\Omega/2\pi=1$\,THz is 
needed),
and a wide gap to reduce the absorption of the generated harmonics.
One option to satisfy these requirements is to use modulated
semiconductor nanowires.  Here we estimate the growth parameters of an
InAs/InP nanowire\cite{bjork:1058}, which would have a near optimal band structure for generating the
third harmonic of a 1\,THz driving field.

\begin{figure}[htbp]
 \centering
\includegraphics[width=3in]{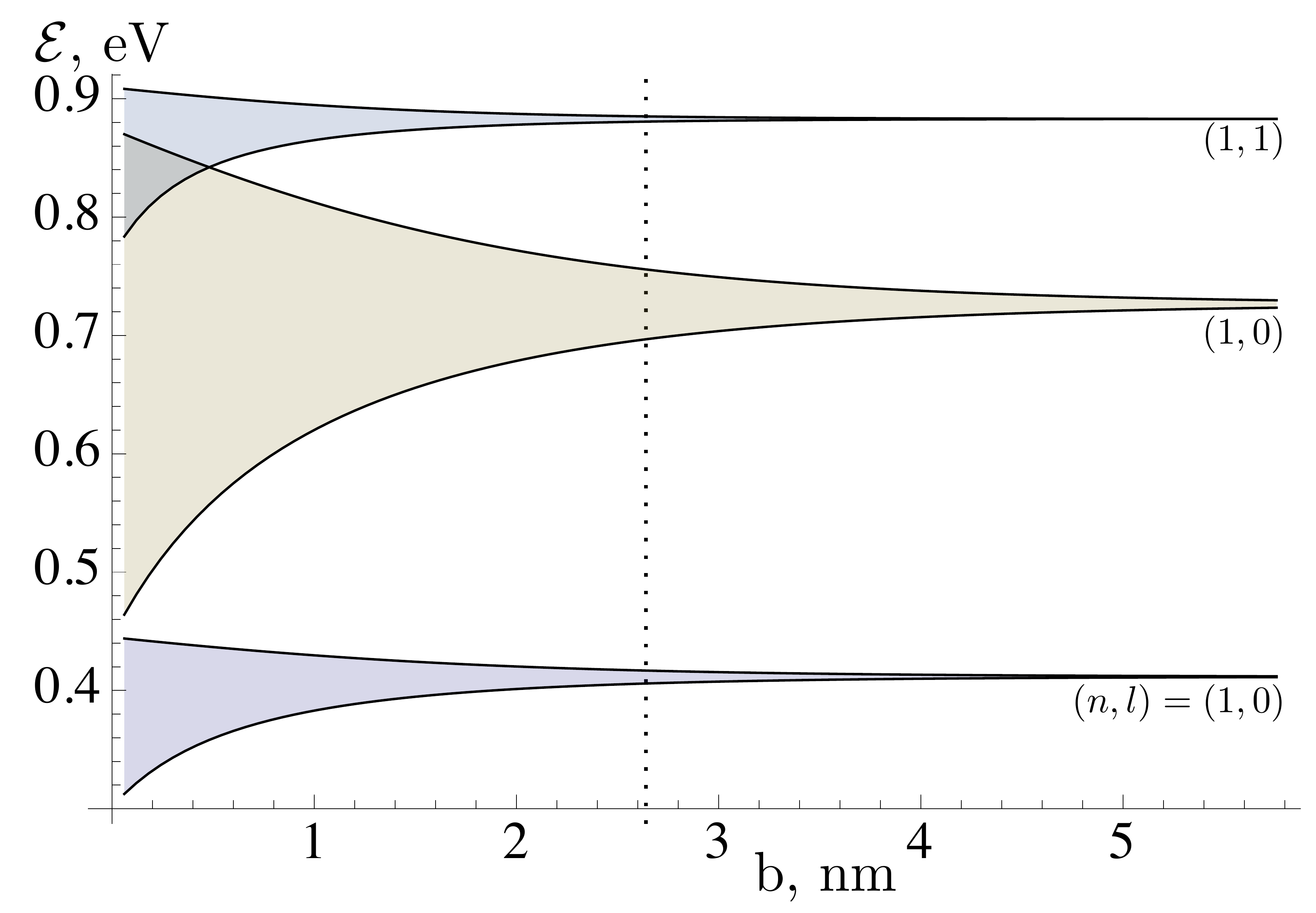}
\caption{Energies of the three lowest bands computed using the
  Kronig-Penney model with effective mass modulation corresponding to
  an InAs/InP nanowire with diameter $d=20$\,nm, InAs well width
  $w=6.0$\,nm, plotted as a function of InP barrier width, ${b}$.  The
  labels indicate the radial $n$ and angular $l$ quantum numbers of
  dimensional quantization.  The dashed line at $b=2.64$\,nm gives a
  bandwidth $10.9$\,meV, or tight-binding parameter $J=2.7$,
  corresponding to the first maximum of the third harmonic in
  Fig.~\ref{fig:I3vJ}.}\label{fig:KPLevJ0R1}
\end{figure}

We calculate the band structure of the modulated nanowire modeling it
as a stack of cylinders with isotropic (bulk) electron effective
masses $m^*_{\rm InAs}=0.073 {m}_{e}$ and $m^*_{\rm InP}=0.027
{m}_{e}$ for the InAs and InP carriers respectively, as
appropriate for the nanowire diameter we
used\cite{Moreira-Venezuela-Miwa-2010}.  We used the barrier height of
$V_0=0.636$\,eV, found from the four-band model simulations, which is
close to experimentally observed\cite{Thelander-2004,bjork:1058}
$0.6$\,eV.  To ensure a relatively large gap, we chose the nanowire
diameter $d=20$\,nm, and InAs well width $w=6.0$\,nm.  Separating 
the radial and angular parts of the
corresponding wave functions, we obtained a version of the
Kronig-Penney model with effective mass modulation, and effective
barrier dependent on the transverse momentum $\hbar\kappa_{nl}$.
We plot the first few allowed energy bads as a function of InP barrier
width $b$ in Fig.~\ref{fig:KPLevJ0R1}.

In particular, we conclude that an InAs/InP nanowire of diameter
$d=20.0$\,nm, well width of $w=6.0$\,nm, and barrier width of
$b=2.64$\,nm [Fig.~\ref{fig:nanowire} (a)] would have the lowest band
with a width of approximately $10.9$\,meV.  The next band would be
separated by a gap of $280$\,meV [Fig.~\ref{fig:KPLevJ0R1}].  These
parameters are near optimal for third harmonic generation at
$\Omega/2\pi=1$\,THz.

One possible device design could involve depositing of a number of
parallel modulated nanowires on a substrate, with an $s$-polarized
driving field incident on the surface at angle $\theta$ so
that the electric field of the wave be directed along the nanowires
[Fig.~\ref{fig:nanowire} (b),(c)].  Then
both the reflected signal and the first harmonic are going to be
propagating at the same reflection angle $\theta$, while the
propagation direction of the third harmonic can be found from the
Snell's law, $\sin\theta = 3\sin\phi$, which accounts for the
wavelengths ratio.

\section{Discussion}

In this work we suggest a possibility that frequency multiplication
due to periodically-driven Bloch oscillation may be possible in a
quasistationary setting, with the help of a narrow-band
one-dimensional conductor.  A quasi-equilibrium electron distribution
is possible because the energy absorbed from the driving field is continuously
dissipated by the bulk phonons.

For a periodically modulated InAs/InP nanowire with the period
$a=8.64$\,nm, and the driving field frequency $\Omega/2\pi=1$\,THz,
the emission of the first harmonic is suppressed with the
dimensionless vector potential amplitude $A_0\,a\approx 3.83$, which
gives the electric field amplitude $E_0=\hbar\Omega \,A_0/e\approx
1.8\times 10^6$\,V/m, corresponding to the energy flux of about
$0.5$\,MWt/cm$^2$.  At this kind of power, many effects could lead to eventual run-away
overheating of the system, e.g., direct
absorption by the substrate, or even a relatively weak disorder
scattering in the nanowire.  We hope that a quasi-continuous operation would
still be possible, with the driving field pulse duration of a few
microseconds, as opposed to few picoseconds in the
experiment\cite{Ghimire-2011}.

\section{Acknowledgements} 
The authors are grateful to Goutam Chattopadhyay, Ken Cooper, and
Robert A.~Suris for multiple helpful discussions, and to Craig Pryor
for letting us use his dot code for nanowire calculations.  This work
was supported in part by the U.S. Army Research Office Grant
No. W911NF-11-1-0027, and by the NSF Grant No. 1018935.

\providecommand*\mcitethebibliography{\thebibliography}
\csname @ifundefined\endcsname{endmcitethebibliography}
  {\let\endmcitethebibliography\endthebibliography}{}

\end{document}